\documentclass[conference]{IEEEtran}
\IEEEoverridecommandlockouts
\usepackage{cite}
\usepackage{amsmath,amssymb,amsfonts}
\usepackage{float}
\usepackage{graphicx,lipsum,multicol}
\usepackage{caption}
\usepackage{subcaption}
\usepackage{mwe}
\usepackage{textcomp}
\usepackage{xcolor}
\usepackage{amsmath}
\usepackage[english]{babel}
\usepackage[utf8]{inputenc}
\usepackage{mathbbol}
\usepackage{amssymb}  
\usepackage{authblk}
\usepackage{stackengine}
\usepackage[linesnumbered,ruled,vlined]{algorithm2e}
\usepackage[noend]{algpseudocode}
\newcommand{\RomanNumeralCaps}[1]{\MakeUppercase{\romannumeral #1}}
\def\BibTeX{{\rm B\kern-.05em{\sc i\kern-.025em b}\kern-.08em
    T\kern-.1667em\lower.7ex\hbox{E}\kern-.125emX}}
\makeatletter
\newcommand*\titleheader[1]{\gdef\@titleheader{#1}}
\AtBeginDocument{%
	\let\st@red@title\@title%
	\def\@title{%
		\bgroup\normalfont\small\centering\@titleheader\par\egroup
		\vskip0.6em\st@red@title}
}
\makeatother

\title{Managing Fog Networks using Reinforcement Learning Based Load Balancing Algorithm}
\titleheader{This is the authors’ version of the paper that has been accepted for publication in IEEE Wireless Communications and Networking Conference,  15- 18 April 2019, Marrakesh, Morocco}

\begin{document}

\author[1]{Jung-yeon Baek}
\author[1]{Georges Kaddoum}
\author[1]{Sahil Garg}
\author[1]{Kuljeet Kaur}
\author[2]{Vivianne Gravel}
\affil[1]{Dept. Electrical Engineering, ETS, University of Quebec, Montréal, QC, Canada}
\affil[ ]{E-mails: jungyeon.baek.1@ens.etsmtl.ca, georges.kaddoum@etsmtl.ca, \protect\\ garg.sahil1990@gmail.com, kuljeet0389@gmail.com}
\affil[2]{B-CITI Inc., 355 rue Peel, Montréal, QC, Canada, H3C 2G9}
\affil[ ]{E-mail: vivianne.gravel@b-citi.com}

\maketitle

\begin{abstract}
The powerful paradigm of Fog computing is currently receiving major interest, as it provides the possibility to integrate virtualized servers into networks and brings cloud service closer to end devices. To support this distributed intelligent platform, Software-Defined Network (SDN) has emerged as a viable network technology in the Fog computing environment. However, uncertainties related to task demands and the different computing capacities of Fog nodes, inquire an effective load balancing algorithm. In this paper, the load balancing problem has been addressed under the constraint of achieving the minimum latency in Fog networks. To handle this problem, a reinforcement learning based decision-making process has been proposed to find the optimal offloading decision with unknown reward and transition functions. The proposed process allows Fog nodes to offload an optimal number of tasks among incoming tasks by selecting an available neighboring Fog node under their respective resource capabilities with the aim to minimize the processing time and the overall overloading probability. Compared with the traditional approaches, the proposed scheme not only simplifies the algorithmic framework without imposing any specific assumption on the network model but also guarantees convergence in polynomial time. The results show that, during average delays, the proposed reinforcement learning-based offloading method achieves significant performance improvements over the variation of service rate and traffic arrival rate. The proposed algorithm achieves 1.17$\%$, 1.02$\%$, and 3.21$\%$ lower overload probability relative to random, least-queue and nearest offloading selection schemes, respectively.

\end{abstract}


\section{Introduction}
Over the last decade, Cloud computing has emerged as an important trend which assisted in moving the computing, control, and data storage resources across the geographically distributed data centers. Today, however, Cloud computing is encountering growing challenges such as-unpredictable high communication latency, privacy gaps and related traffic loads of networks connecting to end devices \cite{b1,b2,b3}. To address some of these limitations, Fog computing has emerged as a promising paradigm that brings the computation closer to the physical IoT devices deployed at the network edge; commonly referred to as ‘Things’ (sensors, mobile phones, edge switches/routers, and vehicles, etc.) \cite{b1,b2,b3,b4,b5}. For example, commercial edge routers having high processing speed and equipped with a large number of cores and communicating with the external or border layer of the network, have the potential to become new servers for Fog networks \cite{b20}. 

Essentially, Fog computing is a highly virtualized platform that offers computing capabilities to allow various applications to run anywhere. To tackle the scalability issues of traditional centralized control architectures, Software-Defined Network (SDN) is the most viable network technology in the Fog environment \cite{b4,b5,b6,b7}. SDN-based Fog architectures provide a centralized controller with global knowledge of the network state which is capable of controlling Fog nodes services while Fog nodes simply accept policies from the controller without understanding various network protocols standards \cite{b22}. In \cite{b21}, the scenario showed that the Fog node includes an SDN controller which handles the programmability of the network edge devices by either a fully distributed or centralized management. 

Although Fog networking is a promising technology to cope with the disadvantages of Cloud and the existing networks, but there are still challenges that remain to be assessed in the future. Most importantly, there is a need for a distributed intelligent platform at the edge that manages distributed computing, networking, and storage resources. In Fog networks, however, making an optimal distribution decision faces a lot of challenges due to uncertainties associated with task demands and resources available at the Fog nodes \cite{b11} and the wide range of computing power capacities of nodes. Furthermore, the distribution decision should also consider the communication delay between nodes, which can lead to prolonged processing time [8]–[10]. Therefore, the challenges being faced by Fog computing paradigm are varied and many; they include crucial decisions about i) whether Fog nodes should be offloaded or not, ii) an optimal number of tasks to be offloaded, and iii) mapping of incoming tasks to available Fog nodes under their respective resource capacities.

The existing schemes proposed in the literature have majorly concentrated on load balancing and cooperative offloading in Fog environment \cite{b8, b9, b10, b11, b12}. However, most of these approaches impose many restrictions and specific assumptions over the networks; which often do not relate with realistic Fog networks \cite{b15}. Additionally, these approaches require non-trivial mathematical equations when more and more restrictions are considered. 

Under the scope of the above challenges, the proposed algorithm formulates the offloading problem as a Markov decision process (MDP) subject to the dynamics of the system in terms of Fog nodes behavior. This problem allows Fog nodes to offload their computation intensive tasks by selecting the most suitable neighboring Fog node in the presence of uncertainties on the task demands and resource availability at the Fog nodes. However, the system cannot precisely predict the transition probabilities and rewards due to dynamically changing incoming task demands and resource status. To solve this problem, this paper uses the classic model-free reinforcement learning algorithm, Q-learning, which can be used to solve MDPs with unknown reward and transition functions by making observations from experience. 

Because of a reinforcement learning methods' advantages, it has been largely discussed in developing load balancing problems. For instance, the authors, in \cite{b17}, implemented reinforcement learning for distributed static load balancing of data-intensive applications in a heterogeneous environment. Dutreilh \textit{et al.} \cite{b18} proposed a learning process for automatic resource allocation in a cloud. Likewise, authors in \cite{b14} proposed a deep reinforcement learning based offloading scheme for a user in an Ad-hoc Mobile Cloud. However, the proposed Q-learning in Fog network differs from existing state-of-the-art in terms of three primary reasons. First, the environment is changed by multiple Fog nodes according to individual incoming requests from end devices. Second, each Fog node may have a different computing capacity which spells out the differences in resource availability between Fog nodes. Third, the distance between Fog nodes affects decision-making for an action which is chosen in a way that minimizes the communication time. 

\subsection{Contributions}
The main objective of the proposed algorithm is to choose optimal offloading decisions while minimizing tasks processing delay and node overload probability. In addition, to provide more distributed and scalability for Fog network model. The novelty of the proposed algorithm is the fact that it accommodates variable incoming task rates and different computing capacities of each Fog node as well as the distance between Fog nodes in its formulation. Moreover, it simplifies the algorithmic framework without any specific assumption on the considered network model, whereas other related works generally impose restrictions on the network so as to simplify non-trivial mathematical equations. Towards this end, the proposed model merges with SDN-based Fog computing where SDN Fog controller directly controls, programs, orchestrates and manages network resources. Further, SDN Fog nodes serve the end users’ request and deliver information based on collected traffic information to the controller \cite{b15}. Besides, the proposed reward function is defined with the aim to minimize the processing time and the overall overloading probability. Since a controller has the global view of the network, it can observe the reward and next state for current state and current action. This advantage can make the proposed Q-learning with guarantees both performance and convergence rates \cite{b15}. As a result, SDN Fog controller can find optimal actions which help nodes to optimally select a neighboring node to which it can offload its requested tasks. Then, nodes also determine how many tasks to offload to a neighboring node chosen based on the size of its task demand and the number of tasks currently remaining in its queue. The key contributions of this paper are listed as follows.
\begin{itemize}
	\item The proposed Q-learning based offloading decision lets the controller decide the optimal actions according to the reward function. This is an attractive feature because the controller has the ability to define their reward function based on the required performance. 
	\item The proposed Q-learning based offloading decision can expect how good the present offloading is over the future, which achieves the exceptional overall system performance. 
	\item The decision-making process using reinforcement learning is on-demand and pay-as-you-go, which makes the proposed method compatible with SDN architecture \cite{b15}. Hence, it can be regarded as a load balancing application running on an SDN Fog controller. These attractive features provide an important opportunity to apply our load balancing problem within SDN Fog networks.
\end{itemize}

\subsection{Organization}
In section 2, this paper discusses load balancing in Fog networks. Section 3 is dedicated to both the system description for the proposed algorithm and the problem formulation. Details on the proposed reinforcement learning-based offloading algorithm is described in Section 4. The simulation results and our analysis of these results are presented in Section 5. Finally, Section 6 concludes this paper and gives some insight on possible future work.
\section{Load balancing in Fog networks}
Load balancing in Fog networks refers to efficiently distributing incoming workload across a group of processing Fog nodes so that the capacity of concurrent users and the reliability of requested tasks increase. It can be categorized under two different methods; static and dynamic load balancing methods \cite{b16}. Static load balancing distributes the workload using prior knowledge of task requests, which is determined at the beginning of the execution. The main drawback of static methods is that the allocation of tasks cannot be changed during the process execution to reflect changes in traffic load. In contrast to the static method, dynamic load balancing allocates tasks dynamically when one of the nodes becomes under-loaded. In other words, it can update the allocation of tasks continuously depending upon the newest knowledge of traffic loads. Hence, the accurate real-time load prediction is necessary for effective load balancing. In virtual machine (VM) environments, multiple VMs share the resources on the same physical machine (PM). Therefore, a load balancing function would be much more complicated in this environment \cite{b7,b12}.

Comparing to traditional servers, Fog node is essentially designated based on the characteristics and features of end devices. Fog nodes are formed by at least one or more physical devices with high processing capabilities \cite{b6}. For a better understanding, a Fog node would be a logical concept, with a heterogeneous type of devices as its physical infrastructure. In this way, the Fog node encompasses end devices together while the processing capacity in these end devices should be presented in terms of virtual computing units. Hence, all physical devices of a Fog node are aggregated as one single logical entity able to seamlessly execute distributed services as if those were on a single device \cite{b6}. In this paper, the Fog node is described as a processing server like a mobile base station but also including the integrated computing capacity of end devices. For example, vehicles and desktop computers can also share their computational power to provide task requests. The Fog node of these end devices can be responsible for managing their resources and also communicate between all Fog nodes. 
\section{system model}
This section fisrtly define the proposed Fog architecture design and the system model. Then, the MDP-based offloading problem formulation are presented. 
\subsection{System description}
In virtualized Fog network, the SDN Fog controller and the SDN Fog nodes are essential infrastructures in the load balancing (Hereinafter, SDN Fog controller and SDN Fog nodes are called controller and nodes, respectively). End users are directly connected to the nodes where they can submit application requests locally. The nodes are connected according to certain network topology, and are logically connected to the controller. In Fig. 1, the paper show a Fog network architecture of the proposed system. This paper considers a dynamic load balancing technique using distributed Fog computing mechanism in which nodes can offload their computation tasks to a neighboring node with available queue spaces in terms of computing capabilities and task demands distributions. Load balancing algorithms are typically based on a load index, which provides a measure of the workload at a node relative to some global average. The load index is used to detect a load imbalance state, in which the load index at one node is much higher or much lower than the load index on the other nodes. The length of the CPU queue has been shown to be remaining resource information that can be used as a good load index on time-shared workstations when the performance measure of interest is the average response time \cite{b14,b18}. 
 
The controller is responsible for information aggregation and decision-making, whereas the nodes work directly to serve the end users and deliver information based on collected traffic information and queue status to controllers. The system works as follows. First, the controller builds a network map based on the queue information delivered from the nodes. Next, the controller runs algorithms to determine whether the present node should offload its requested tasks or not and if so, how many tasks to offload to a neighboring node chosen based on the size of its task demand and the number of tasks currently remaining in its queue. The main objective is to choose optimal offloading actions while minimizing tasks processing delay and node overload probability.

\begin{figure}[t]
	\centering
	\includegraphics[width=\linewidth]{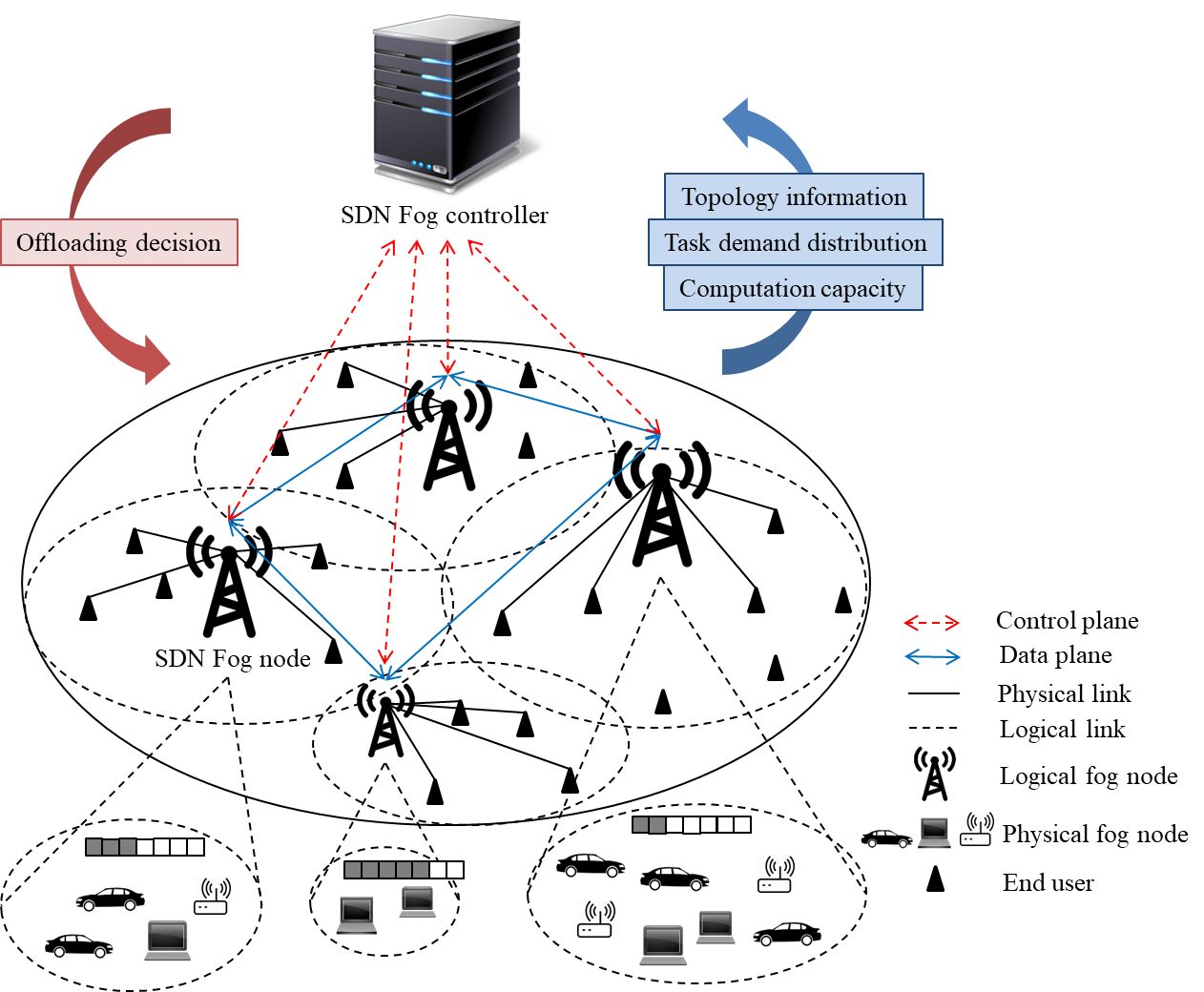}
	\caption{The system model for the SDN-based Fog architecture.}
	\label{figure1}
\end{figure}

\subsection{Problem formulation} 
To achieve the required performance, the proposed load balancing problem is formulated as an MDP for which this paper proposes an algorithm with performance guarantees. MDP involves a decision agent that repeatedly observes the current state $s$ of the controlled system, takes a decision $a$ among the ones allowed in that state (${a}\in{A(s)}$) and then observes a transition to a new state $s'$ and a reward $r$ that will affect its following decisions. MDP is a stochastic rule by which the agent selects actions as a function of states. Hence, the new state and the reward observed from the transition probability distributions can fully characterize the behavior of the underlying controlled system. In this system, the controller selects actions as a function of present states, taking into account the following states and the rewards observed from all nodes.

In particular, our MDP is characterized by a 4-tuple $\big \langle S,A,P,R \big \rangle $, detailed as below:

\begin{itemize}
	\item $S=\{s=$($n^l,w,Q$)$\}$ is the state space
	
	\begin{itemize}
		\item ${n^l}\in{\mathbb{N}}$ ($1 \le n^l \le N$) is the node which has the requested tasks to be allocated from end users
		\item ${w}\in{\mathbb{N}}$ ($1 \le w \le {W_{max}}$) is the number of tasks to be allocated per unit time
		\item $Q=\{$($Q_1,...,Q_N$)$|Q_i\in$\{$0,1,...,Q_{i,max}$\}$\}$ is the number of tasks currently remaining in the nodes' queue
	\end{itemize}
	
	\item $A=\{a=$($n^o,w^o$)$\}$ is the action space
	
	\begin{itemize}
		\item ${n^o}\in{\mathbb{N}}$ ($1 \le n^o \le N,n^o\neq n^l$) is defined as a neighboring node within the considered Fog network and that is being offloaded by the node $n^l$.
		\item ${w^o}\in{\mathbb{N}}$ ($1 \le w^o \le {W_{max}}$) is the number of tasks to be offloaded to a neighboring fog node $n^o$.
		 Let $A(s) \subseteq A$ be the set of actions that can be taken at the state $s$. 
		 $A(s)$ is determined such that the node $n^l$ can only offload the tasks to another node with equal or less number of tasks currently requested. 
		 Depending on an action $a$, the number of tasks to be locally processed ($w^l$) is decided with regard to the available queue space of the node $n^l$. 
	\end{itemize}

	\item $P : S \times A \times S \rightarrow [0,1] $ is the transition probability distribution $P(s'|s,a)$ of a new state $s'$ given that the system is in state $s$ and action $a$ is chosen.
	
	\item $R : S \times A \rightarrow \mathbb{R}$ is the reward when the system is in state $s$ and action $a$ is taken. The main goal of the system is to make an optimal offloading action at each system with the objective of maximizing the utility while minimizing the processing delay and overload probability. Hence, the proposed system defines the immediate reward function $R(s,a)$ given an action $a$ at state $s$ as follows. 
	\begin{equation} \label{one}
	R(s,a) = U(s,a) - (D(s,a) + O(s,a)),
	\end{equation}
	where $U(s,a)$, $D(s,a)$ and $O(s,a)$ represent the immediate utility, immediate delay and overload probability function, respectively.
	
	\begin{itemize}
		\item The immediate utility $U(s,a)$ is calculated as
		\begin{equation} \label{two}
		U(s,a) = r_u\log (1+w^l+w^o),
		\end{equation}
		where, $r_u$ is a utility reward.
		
		\item The immediate delay $D(s,a)$ is calculated as
		\begin{equation} \label{three}
		D(s,a) = \chi_d\cdot\frac{t^w+t^c+t^e}{(w^l+w^o)},
		\end{equation}
		
		where, $\chi_d$ is a delay weight,
		
		\begin{enumerate}
			\item The average waiting time $t^w$ at the queue of the fog node $n^l$ and the offloading fog node $n^o$ is 
			\begin{equation} \label{four}
			t^w = \frac{Q^l}{\overline{\mu ^l}}\mathbb{1}(w^l\neq 0)+\Big(\frac{Q^l}{\overline{\mu ^l}}+\frac{Q^o}{\overline{\mu ^o}}\Big)\mathbb{1}(w^o\neq 0),
			\end{equation}
			where, $\mu ^i$ is the computing service rate of node $n^i$.
			
			\item The communication time of task offloading $t^c$ is
			\begin{equation} \label{five}
			t^c = \frac{2\cdot T\cdot w^o}{r_{l,o}},
			\end{equation}
			where, $T$ is the data size of a task, $r_{l,o}$ is the fog transmission service rate from $n^l$ to $n^o$ given by:
			\begin{equation} \label{six}
			r_{i,j} = B\cdot\log \Big(1+\frac{g_{i,j}\cdot P_{t_{x},i}}{B\cdot N_0}\Big),
			\end{equation}
			where, $B$ is the bandwidth per a node, $g_{i,j}\triangleq \beta_1 {d_{i,j}}^{-\beta_2}$ is the channel gain between nodes $n^i$ and $n^j$ with $d_{i,j}$, $\beta_1$, and $\beta_2$ being the distance between two nodes, the path loss constant and path loss exponent, respectively. The variable $P_{t_{x},i}$ denotes the transmission power of node $n^i$ and $N_0$ is the noise power spectral density, which is defined at normalized thermal noise power -174 dBm/Hz.
			
			\item The execution time $t^e$ by the fog node $n^l$ and the offloaded fog node $n^o$ is
			\begin{equation} \label{seven}
			t^e = \frac{I\cdot CPI \cdot w^l}{f^l}+\frac{I\cdot CPI \cdot w^o}{f^o},
			\end{equation}
			where, $I$ is the number of instructions per task, $CPI$ is CPU cycles per instruction and $f^i$ is the CPU speeds of a node $n^i$.
			
		\end{enumerate}
		
		\item The overloaded probability $O(s,a)$ is calculated as
		\begin{equation} \label{eight}
		O(s,a)=\chi_o\cdot\frac{w^l\cdot P_{overload,l}+w^o\cdot P_{overload,o}}{w^l+w^o},
		\end{equation}
		\begin{equation} 
		P_{overload,i} =\frac{\max(0,\lambda_i-(Q_{i,max}-Q'_i))}{\lambda_i},
		\end{equation}
		\begin{equation} 
		Q'_i = \min(\max(0,Q_i-\overline{\mu^i})+w^i,Q_{i,max}),
		\end{equation}
		where, $\chi_o$ is an overload weight and $\lambda_i$ is the task arrival rate arriving at node $n^i$ which can be modeled by a Poisson process. In (10), $Q'_i$ represents the next estimated queue state of node $n^i$ in state $s$ and action $a$ is taken.
		
	\end{itemize}

\end{itemize}
With the transition probability $P$ and reward $R$ being determined prior to the execution of the controlled system, the MDP can thus be solved through traditional dynamic programming (DP) algorithms. The key idea of DP is the use of value functions to organize and structure the search for good decisions. The optimal action at each state is defined as the action that gives the maximum long-term reward, which is the discounted sum of the expected immediate rewards of all future decisions about state-action starting from the current state. An immediate reward received $k$ time steps further in the future is worth only $\gamma^{k-1}$ times what it would be worth if it were received immediately where $\gamma$ is called a \textit{discount factor}(0$<$$\gamma$$<$1) \cite{b13}. The optimal value function is defined, which satisfy the Bellman optimality equations:
\begin{equation}
\begin{split}
v^*(s) &= \max_{a} \mathbb{E}(R_{t+1}+\gamma v^*(S_{t+1})|S_t=s,A_t=a) \\
& = \max_{a}\sum_{s',r}p(s',r|s,a)[r+\gamma v^*(s')].
\end{split}
\end{equation}

\section{The proposed Reinforcement learning-based offloading algorithm}
In this Section, the reinforcement learning-based offloading algorithm is proposed to address the limitations of the traditional approaches.

As defined in the previous section, in the case where the system can have transition probability functions $P$ and reward $R$ for any state-action pair, the MDP can be solved through DP methods. However, for most cases, the system cannot precisely predict $P$, and $R$; and the system may change the transition probability distributions or rewards. To address these limitations, reinforcement learning is proposed. In reinforcement learning, the lack of information is solved by making observations from experience. Classical DP algorithms are of limited features in reinforcement learning both because of their assumption of a perfect model and because of their great computational cost \cite{b13}. Among the different reinforcement learning techniques, Q-learning is the classic model-free algorithm. It is usually used to find the optimal state-action policy for any MDP without an underlying policy. Given the controlled system, the learning controller repeatedly observes the current state $s$, takes action $a$, and then a transition occurs, and it observes the new state $s'$ and the reward $r$. From these observations, it can update its estimation of the Q-function for state $s$ and action $a$ as follows:
\begin{equation}
Q(s,a) \leftarrow (1-\alpha)Q(s,a)+\alpha \Big[R(s,a)+\gamma\max_{a'\in A_{s'}}Q(s',a')\Big],
\end{equation}
where, $\alpha$ is the \textit{learning rate} (0$<$$\alpha$$<$1), balancing the weight of what has already been learned with the weight of the new observation.

The simplest action selection rule is to select one of the actions with the highest estimated value, i.e., Greedy selection ($a_t\doteq\arg\max_{a}Q_t(a)$).  Thus, the greedy action selection always exploits current knowledge to maximize immediate reward. An important element of Q-learning is the $\epsilon$-greedy algorithm. This latter behaves greedily most of the time, but with small probability, $\epsilon$, select randomly from all the available actions with equal probabilities, independently. Reinforcement learning call this greedy selection and the $\epsilon$ probability of random selection as exploitation and exploration policies, respectively. Exploitation is the right action to do to maximize the expected reward on the one step, whereas exploration may produce the greater total reward in the long term. An advantage of $\epsilon$-greedy algorithm is that, as the number of steps increases, every action will be visited an infinite number of times, thus ensuring that $Q(s,a)$ converges to the optimal value.

The appropriate reward function that the proposed system have chosen is calculated using Eq.(1). The approximate next state $s'$ is obtained after defining its three components mentioned in the Section \RomanNumeralCaps{3}-B; While the next queue state is a deterministic entity,  the next fog node which has task to be allocated and the size of tasks following each node tasks arrival are of stochastic nature and therefore the proposed system models them using Poisson random variables.

The procedures of the proposed Q-learning algorithm is presented in Algorithm 1.

\begin{algorithm}
	\DontPrintSemicolon
	\LinesNumbered
	\textbf{Input} learning rate ($\alpha$), discount factor ($\gamma$), exploration policy ($\epsilon$), service rate ($\mu$), task arrival rate ($\lambda$), and distance vector ($\mathcal{D}$)
	
	\textbf{Output} Optimized offloading table ($\mathcal{Q}$)
	
	\textbf{Set} $Q(s,a)$ := 0 ($\forall$$s \in S$) ($\forall$$a \in A(a)$), \textit{iter} := 0, and $s$ := (1,1,$\{$($Q_1,...,Q_N$)$|Q_i=0\}$)
	
	\While{(iter $\le$ maximum iteration)}{
		 Choose $a \in A(a)$ using $\epsilon$-greedy algorithm\;
		Offload the tasks according to action $a$ and observe next state $s'$ and reward $r$\;
		$Q(s,a)\leftarrow (1-\alpha)Q(s,a)+\alpha\Big[R(s,a)+\gamma\max_{a'\in A_{s'}}Q(s',a')\Big]$\;
		$s \leftarrow s'$
		
		\textit{iter}$\leftarrow$\textit{iter}$+1$
	}
	\caption{The proposed Q-learning algorithm\label{QR}}
\end{algorithm}

\section{Performance evaluation}
\subsection{Simulation settings}
This Section analyzes the performance of the proposed load balancing algorithm in simulations in which the proposed system considers a fog network consisting of $N$ nodes. The system sets a network area of $100\times 100$ $m^{2}$ where nodes were randomly allocated. At the beginning of the Q-learning, since the first Q-value is zero, the algorithm encourages exploration more. That is why the optimal action selection worked for $\epsilon$ = 0.9 and 0.7 between the initial iteration and the last iteration. In addition, utility reward $r_u$ is set to 10. 
The performance of the proposed Q-learning algorithm was compared to some of the existing offloading methods. Specifically, the simulation used least-queue, nearest, and random node selection offloading methods as benchmarks. Least-queue is a classic offloading method, which is implemented by offloading tasks always to the node with minimum queue status. Nearest node selection is an offloading algorithm widely used in IoT, device-to-device communications since the node selects the most adjacent neighboring node aiming to minimize communication delay and energy consumption.

For different evaluation scenarios, the simulation was set to most parameters to default values and vary the exclusively following parameters. 1) task arrival process of nodes; 2) computing service rate of nodes. Over the variation of the task arrival rate, the simulation was considered the average computing service rate of fog node as 1.8. Likewise, over the variation of the computing service rate, the average task arrival rate was kept to 5.2. The detailed simulation parameters are given in Table \ref{table1}.

\begin{table}[h]
	\caption{Parameter values in simulations.}
	\label{table1}
	\centering
	\begin{tabular}{c|c|c}
		\hline\noalign{\smallskip}
		\multicolumn{2}{c}{\textbf{Parameter}} & \textbf{Value}\\
		\hline
		\hline
		\multicolumn{2}{c}{Number of fog nodes ($N$)} & 5\\
		\hline
		\multicolumn{2}{c}{Maximum CPU queue size ($Q_{max}$)} & 10\\
		\hline
		\multicolumn{2}{c}{Learning rate ($\alpha$)} & 0.5\\
		\hline
		\multicolumn{2}{c}{Discount factor ($\gamma$)} & 0.5\\
		\hline
		\multicolumn{2}{c}{Data size per a task ($T$)} & 500 Mbytes\\
		\hline
		\multicolumn{2}{c}{Number of instructions per a task ($I$)} & 200 $\times 10^6$\\
		\hline
		\multicolumn{2}{c}{Number of cycles per a an instruction ($CPI$)} & 5\\
		\hline
		\multicolumn{2}{c}{System bandwidth per node ($B$)} & 2 MHz\\
		\hline
		\multicolumn{2}{c}{Path loss parameter ($\beta_1$,$\beta_2$)} & $(10^{-3},4)$\\
		\hline
		\multicolumn{2}{c}{Transmission power of fog node ($P_{tx,i}$)} & 20 dBm\\
		\hline
		\multicolumn{2}{c}{Weights ($\chi_d$,$\chi_o$)} & $(1, 150)$\\
		\hline
	\end{tabular}
\end{table}


\subsection{Performance analysis}
Simulation results are presented in Fig. 2, 3, and 4 under different system configurations. Fig. 2(a) and 2(b) show the average reward over the variation of the task arrival rate and the computing service rate, respectively. Throughout the different configurations, the proposed load balancing algorithm using reinforcement learning shows higher average rewards than the three other classic load balancing methods. The main reason is that the proposed offloading decision is to offload one of the actions with the highest Q-value. This algorithm implies that the Q-learning based offloading decision minimize processing time and overload probability according to the proposed reward function. Furthermore, the Q-learning can expect how good the present offloading is over the future, which achieves the exceptional overall system performance. The average reward greatly increases as the computing service rate is increased since more number of tasks are executed. Meanwhile, the average reward consistently decreases as the arrival rate is increased because the fog nodes have a relatively high number of tasks in their queues; therefore a lower number of tasks can be allocated by them.

Fig. 3(a) and 3(b) show the average delay over the variation of the task arrival rate and the computing service rate, respectively. The proposed algorithm achieves the minimum average delay. The main reason is that the proposed offloading algorithm accommodates both the node’s queue status and the distance between fog nodes in its formulation. On the other hand, when the computing service rate increases from 3 to 7, the average delay hardly decreases. This result indicates that once a certain level of computing service rate is achieved, for a fixed arrival rate, the delay that is occurring is caused by the communication latency which is defined as the transmission service rate from one fog node to another one offloaded. 

In addition, the average overload probability was observed as shown in Fig. 4(a) and 4(b). The results show that the proposed algorithm greatly reduces the overload probability. Specifically, when the task arrival rate is 9, the proposed algorithm offers 5.77$\%$, 5.17$\%$, and 6.23$\%$ lower overload probability than random, least-queue, and nearest offloading methods. These results highlight the fact that when the node makes an offloading decision with the proposed algorithm, it considers not only nodes’ queue states but also their task arrival distribution. Therefore, the proposed algorithm can minimize the failed allocation and with that the risk that a task can be lost if it arrives when the nodes’ queue is already full. These improvements can be attributed to the Q-learning based offloading decision in accordance with different service rate and task arrival rate of nodes.
\begin{figure}[t]
	\centering
	\includegraphics[width=0.9\linewidth]{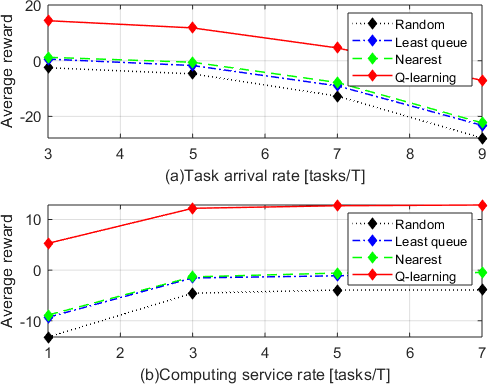}
	\caption{Average reward.}
	\label{figure2}
\end{figure}
\begin{figure}[t]
	\centering
	\includegraphics[width=0.9\linewidth]{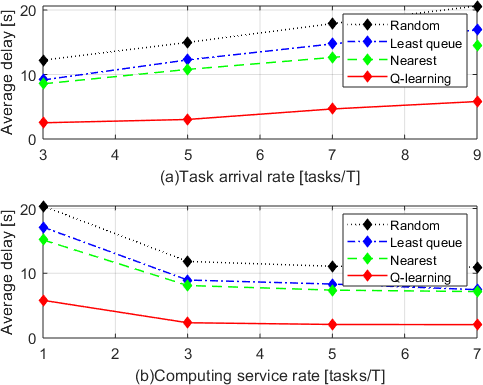}
	\caption{Average delay.}
	\label{figure3}
\end{figure}

\begin{figure}[t]
	\centering
	\includegraphics[width=0.9\linewidth]{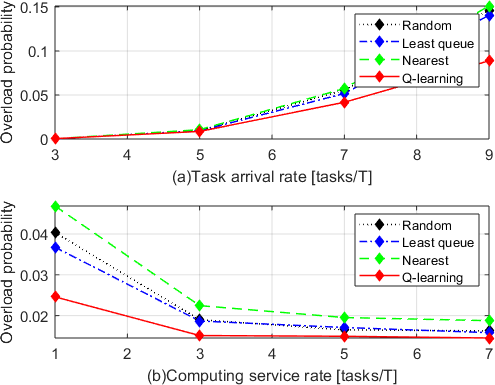}
	\caption{Average overload probability.}
	\label{figure4}
\end{figure}

\section{Conclusion}
In this paper, the Fog dynamic load balancing algorithm has been proposed using a reinforcement learning technique. Load balancing in Fog networks faces a lot of challenges due to uncertainties related to workloads and the wide range of computing power capacities of nodes. The proposed process formulates a Markov decision process to find optimal actions which help nodes to select optimally a neighboring node to which it can offload its requested tasks. Then, nodes also determine how many tasks to offload to a neighboring node chosen based on the size of its task demand and the number of tasks currently remaining in its queue. Actions considered to be optimal if they allow minimal processing time and a minimum overall overload probability. To guarantee the optimal action-selection, the proposed algorithm applied Q-learning with $\epsilon$-greedy algorithm. The performance of the proposed load balancing algorithm is evaluated in different configurations in which nodes offload to their neighboring nodes within a realistic network. The simulation results show that the proposed algorithm can achieve lower average processing delay and lower failed allocation probability due to overloading as compared to existing methods. As the future works, the proposed model-free learning merged with the model-based learning will be considered to experiment, which may give less dependency to exploration policy and bias-free results.

\section*{Acknowledgement}
This research was supported by the NSERC B-CITI CRDPJ 501617-16 grant

\end{document}